\documentclass{emulateapj}

\begin{document}

\slugcomment{Accepted for publication in PASP, Sept 2004 issue}

\title{The Cryogenic Refractive Indices of S-FTM16, a Unique Optical Glass 
for Near-Infrared Instruments}

\author{Warren R.\ Brown}

\affil{Harvard-Smithsonian Center for Astrophysics, 60 Garden St,
Cambridge, MA 02138}

\author{Harland W.\ Epps}

\affil{University of California Observatories/Lick Observatory, Santa
Cruz, CA 95064}

\and

\author{Daniel G.\ Fabricant}

\affil{Smithsonian Astrophysical Observatory, 60 Garden St, Cambridge, MA
02138}

\shorttitle{The Thermal Dependence of S-FTM16's Refractive Index}
\shortauthors{Brown et al.}

\begin{abstract}

	The Ohara glass \hbox{S-FTM16} is of considerable interest for
near-infrared optical designs because it transmits well through the $K$
band and because negative \hbox{S-FTM16} elements can be used to
accurately achromatize positive calcium fluoride elements in refractive
collimators and cameras.  Glass manufacturers have sophisticated equipment
to measure the refractive index at room temperature, but cannot typically
measure the refractive index at cryogenic temperatures.  Near-infrared
optics, however, are operated at cryogenic temperatures to reduce thermal
background.  Thus we need to know the temperature dependence of
\hbox{S-FTM16's} refractive index.  We report here our measurements of the
thermal dependence of \hbox{S-FTM16's} refractive index between room
temperature and $\sim$77 K.  Within our measurement errors we find no
evidence for a wavelength dependence or a nonlinear temperature term so
our series of measurements can be reduced to a single number.  We find
that $\Delta n_{abs}/\Delta T=-2.4\times10^{-6}$ K$^{-1}$ between 298 K
and $\sim$77 K and in the wavelength range 0.6 \micron\ to 2.6 \micron.  
We estimate that the systematic error (which dominates the measurement
error) in our measurement is 10\%, sufficiently low for most purposes.  
We also find the integrated linear thermal expansion of \hbox{S-FTM16}
between 298 K and 77 K is -0.00167 \hbox{m m$^{-1}$}.

\end{abstract}


\section{INTRODUCTION}

	Optical designs of near-infrared (1.0 - 2.4 \micron) instruments
are strongly constrained by a limited selection of near-infrared optical
materials.  Low-index materials such as CaF$_2$, BaF$_2$, LiF, NaCl, and
infrared grade fused quartz are commonly used for windows and simple
achromatic doublets.  However, these low-index materials have a restricted
range of dispersive power that prevents good chromatic correction in 
complex, high-performance optical designs.  In addition, NaCl, LiF,
and BaF$_2$ are hygroscopic and prone to damage.  Zinc Selenide (ZnSe) is
another material suitable for near-infrared lenses, but its high index and
large dispersive power limits its usefulness.  In addition, the high index
of ZnSe makes it difficult to design broadband anti-reflection coatings.

	The internal transmission of most optical glasses with {\it
intermediate} indices and dispersive powers is typically poor at
near-infrared wavelengths, but the internal transmission of the glass
\hbox{S-FTM16} is a fortunate exception.  Figure \ref{fig:intrans} shows
that a 10-mm thick piece of \hbox{S-FTM16} has an internal transmission of
98.7\% at 2.0 \micron\ and 95.3\% at 2.4 \micron.  The dispersive power of
\hbox{S-FTM16} is $(dn/d\lambda)/(n-1)=-0.036155$ at 1.0 \micron\ and
$=-0.026242$ at 2.0 \micron.  To compare this with other infrared
materials, we refer the reader to Table 1 of \citet{epps02}, which sorts
twenty four infrared materials by dispersive power.  The intermediate
dispersive power of \hbox{S-FTM16} is quite similar to the Schott glass
TIFN5 and somewhat similar to the older Schott glass IRG7.  However, TIFN5
and IRG7 are no longer available as stock glasses.

	The intermediate dispersive power of \hbox{S-FTM16} is extremely
valuable for high-performance, near-infrared optical designs.  For
example, \hbox{S-FTM16} is a critical part of the optical designs of
FLAMINGOS-2, a near-infrared multi-slit spectrograph and imager for Gemini
South \citep{epps02}, and MMIRS, an instrument based on FLAMINGOS-2 for
the f/5 MMT and Magellan telescopes \citep{mcleod04}.  Near-infrared
astronomical instruments like FLAMINGOS-2 and MMIRS operate at liquid
nitrogen temperatures (77 K) in order to reduce thermal background.  The
FLAMINGOS-2 and MMIRS optical designs are sensitive to index uncertainties
of a few parts in 10$^{-5}$.  Thus we need to know the refractive index,
$n$, of \hbox{S-FTM16} at cryogenic temperatures near $\sim$77 K and at
near-infrared wavelengths to be sure that the optics of FLAMINGOS-2 and
MMIRS will perform properly.

	Two large S-FTM16 blanks were purchased from Ohara Corporation for
the FLAMINGOS-2 spectrograph optics.  We had a small prism made from one
of the blanks, and we sent the prism to Ohara to obtain melt indices. On
29 August 2003, Ohara obtained index measurements relative to air at room
temperature (25 $^{\circ}$C) and at sea level (760 torr).  Measurement
accuracy was stated to be $< \pm1\times10^{-5}$. Table \ref{tab:ohara}
presents the 12 optical/near-infrared lines we obtained.  We were unable,
however, to obtain cryogenic index measurements commercially.

 \centerline{ \includegraphics[width=2.5in]{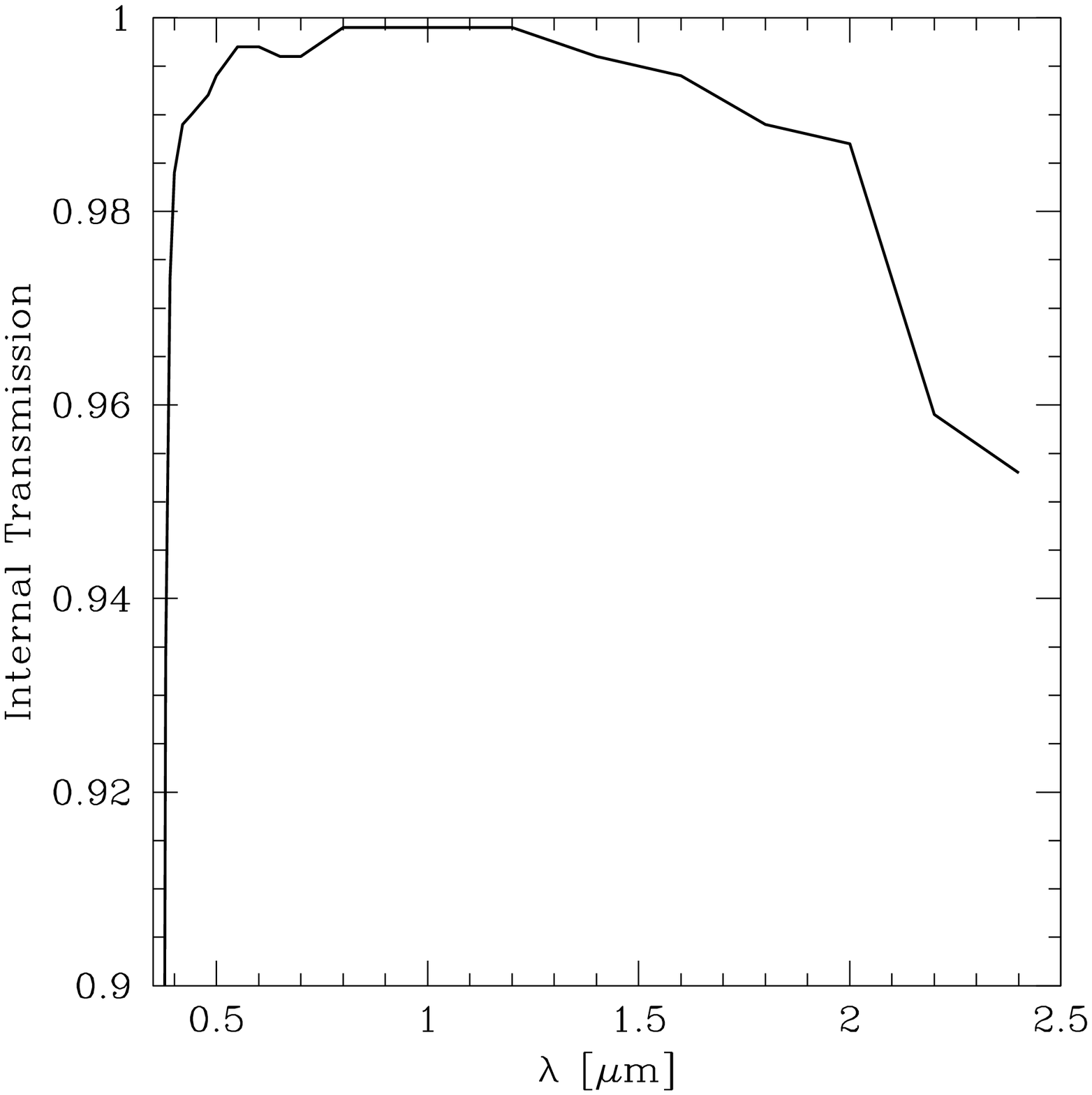} }
 \figcaption{ \label{fig:intrans}
	Internal transmission of a 10-mm thick piece of the
intermediate-dispersion glass \hbox{S-FTM16}.  (Values published by Ohara
Corporation.)}

	~

	Here, we report our measurements of $\Delta n/\Delta T$, the
change in refractive index for \hbox{S-FTM16} between room temperature and
$\sim$77 K.  We will determine \hbox{S-FTM16's} cryogenic index by
converting its melt indices in air to absolute indices in vacuum, and then
adding our value of absolute $\Delta n/\Delta T$.

	Physically, the temperature dependence of the refractive index
depends on a material's density and its electrical properties
\citep{frohlich49}.  Materials expand and contract with temperature, thus
density is temperature dependent.  This is quantified by a material's
coefficient of thermal expansion (CTE), which is wavelength independent.  
Electrically, a material can be neutral while its constituent atoms and
ions have dipole moments that cause a macroscopic polarizability.  The
refractive index and dielectric constant depend on this macroscopic
polarizability.  The macroscopic polarizability is wavelength dependent.  
Because the ordering of microscopic dipole moments in a material depends
on temperature, the macroscopic polarizability is also temperature
dependent.  Generally speaking, microscopic dipole moments will experience
the most re-ordering near the melting point, and the least re-ordering
near 0 K.  This suggests we may not see a strong contribution from the
macroscopic polarizability in our temperature regime.  Interestingly, the
refractive index changes due to density and polarizability are usually
opposite in sign. Refractive index changes due to density tend to produce
a negative $dn/dT$, and those due to polarizability tend to produce a
positive $dn/dT$ \citep{tropf1995}.  The balance of the two effects
determines the final magnitude and sign of $dn/dT$.


	The Ohara catalog contains thermo-optical measurements for
\hbox{S-FTM16}, but for the limited temperature range -30 $^{\circ}$C to
+70 $^{\circ}$C and for the limited wavelength range 0.435835 \micron\ to
1.01398 \micron.  Table \ref{tab:oharadndt} summarizes the thermo-optical
coefficients for \hbox{S-FTM16} presented in the Ohara catalog, converted
to an absolute $dn/dT$.  Absolute $dn/dT$ refers to a material with a
vacuum/glass interface as opposed to a relative $dn/dT$ which refers to an
air/glass interface. Throughout this paper we discuss absolute $dn/dT$, or
$dn_{abs}/dT$. It is apparent from Table \ref{tab:oharadndt} that
$dn_{abs}/dT$ of \hbox{S-FTM16} varies with wavelength and 

\centerline{ \includegraphics[width=3.7in]{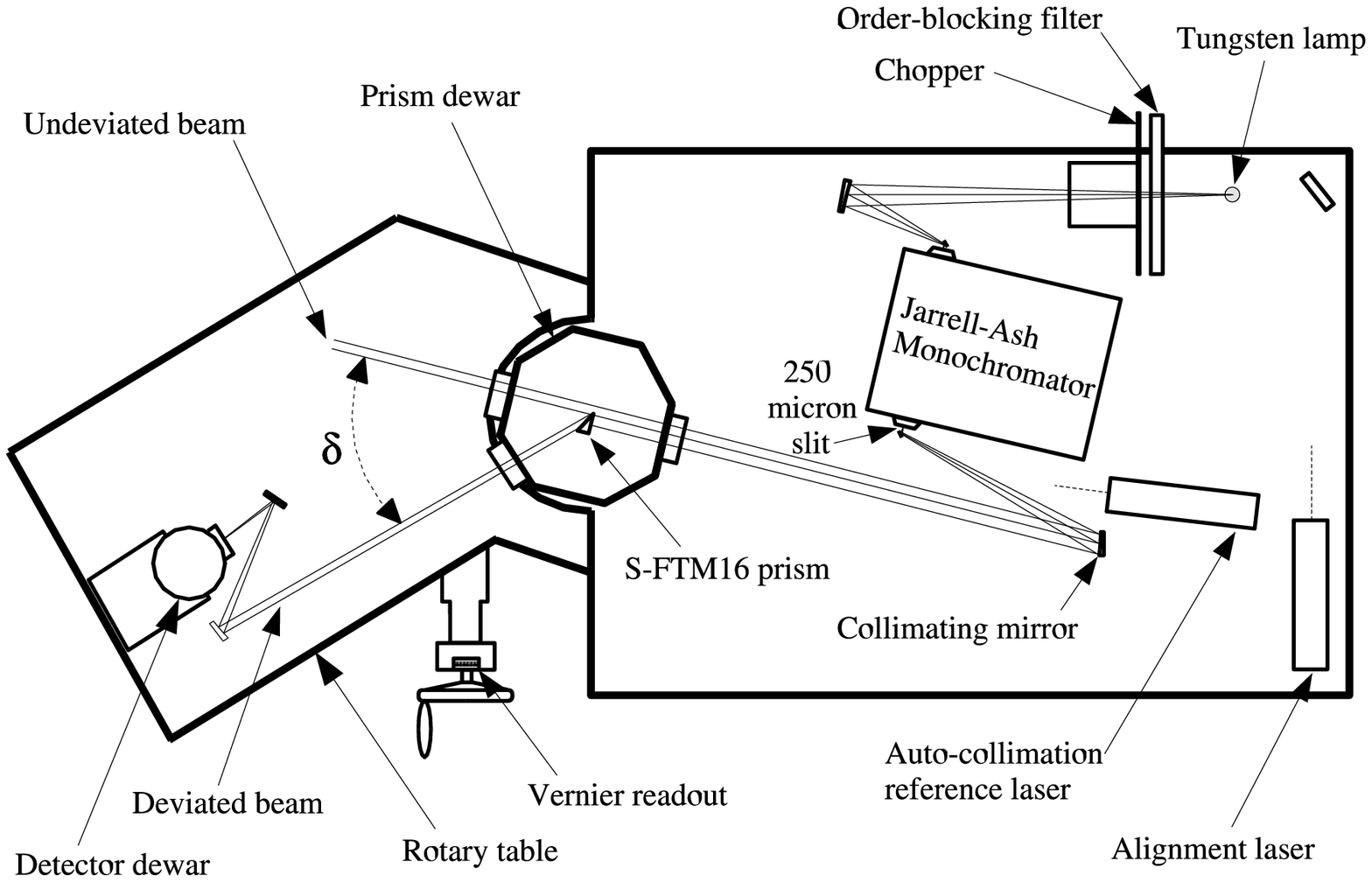} }
 \figcaption{ \label{fig:cryorefrac}
	The layout of the Cryogenic Refractometer.  The refractive index
of the \hbox{S-FTM16} prism at a given temperature and wavelength is found
by measuring the angle $\delta$ between the deviated and undeviated
beams.  Figure based on a drawing by J.\ Palmer.}

	~

\noindent temperature.  We note that the dependence of $dn_{abs}/dT$ with
wavelength and temperature decreases as the temperature drops and the
wavelength increases, trends that are favorable for our application.

	The paper is organized as follows.  We begin in \S
\ref{sec:cryorefrac} by describing the experimental apparatus we used to
measure \hbox{S-FTM16's} cryogenic refractive indices.  We discuss the
experimental data and errors in \S \ref{sec:data}.  We then present our
measurements of \hbox{S-FTM16's} refractive index, thermo-optical
coefficients, and cryogenic CTE in \S \ref{sec:results}.  We conclude in
\S \ref{sec:conclusions}.

\section{THE CRYOGENIC REFRACTOMETER} \label{sec:cryorefrac}

	We used the Cryogenic Refractometer at the University of Arizona
Optical Sciences Center to measure the change of the \hbox{S-FTM16}
refractive index between room and cryogenic temperatures.  The Cryogenic
Refractometer is an instrument built by J.\ Palmer and collaborators
\citep{wolfe80} that uses the modified minimum deviation method of
\citet{platt75} to measure indices of refraction (see Figure
\ref{fig:cryorefrac}).  The advantage of the modified minimum deviation
method is that it allows a prism to be mounted in a fixed cryogenic dewar.  
The disadvantage of the modified minimum deviation method is that it
provides a less precise refractive index measurement compared to the more
traditional minimum deviation and Littrow methods (because the collimated
beam of light makes only a single pass through our prism).  Optical glass
manufacturers, using traditional methods, can provide index measurements
with precisions as good as $\pm1\times10^{-6}$ at room temperature.  By
comparison, our measurements have precisions of $\pm3.5\times10^{-5}$ at
cryogenic temperatures.

	The layout of the Cryogenic Refractometer is shown in Figure
\ref{fig:cryorefrac}.  The refractometer measures the deviation angle,
$\delta$, of a collimated beam of light of wavelength $\lambda$ that
passes through a prism at temperature $T$.  The prism's refractive index
$n(\lambda, T)$ is found from
	\begin{equation} n(\lambda, T) = \frac{\sin{(\alpha +
\delta(\lambda, T)})}{\sin{\alpha}} \label{eqn:n} \end{equation} 
	where $\alpha$ is the apex angle of the prism and $\delta$ is the
angle between the undeviated and deviated beams of light.

	The test prism was mounted in a copper fixture inside the dewar.
Temperatures between 60 K and 77 K were obtained by pumping on the LN$_2$
reservoir; temperatures between 77 K and 100 K were obtained with a heater
adjacent to the prism.  Temperatures were measured with two silicon 
diodes.  One diode was located between the prism and the heater and the 
other was located on the copper fixture below the prism.

	The refractometer uses a LN$_2$-cooled HgCdTe detector for 1.1 -
2.6 \micron\ measurements and a room-temperature Si detector for 0.5 - 1.2
\micron\ measurements.  We measured the angle $\delta$ by rotating the
detector on a precision rotary turntable (Soci\'{e}t\'{e} Genevoise PI-4)
to locate the undeviated and deviated beams. The collimated beam was large
enough to pass both through the prism (deviated) and around the prism
(undeviated).  We used a chopper and lock-in amplifier to improve the
signal-to-noise of the beams.

	We used a tungsten ribbon-filament lamp as the light source for
all of our measurements.  A 300 line mm$^{-1}$ grating inside a Jarrell
Ash Model 27 monochromator selected the desired wavelength.  We used the
300 line mm$^{-1}$ grating in first, second, and third orders to cover our
full wavelength range.  A 250 \micron\ exit slit provided a spectral
bandwidth ranging from 3 to 1 nm FWHM, depending on the grating order.

\section{DATA AND ERRORS} \label{sec:data}

\subsection{Measurements}

	We measured the refractive index as follows.  First, we allowed
the prism temperature to stabilize.  We then auto-collimated the prism to
the collimated beam of light.  We selected the desired wavelength with the
Jarrell Ash monochrometer, using an order blocking filter as necessary.  
Finally, we rotated the arm holding the detector and located the
undeviated and deviated beams.  We repeated the last two steps until all
wavelengths at a given temperature were measured.

	The spatial profile of the collimated beam at the detector was the
convolution of two narrow rectangle slits and so appeared approximately
Gaussian in shape.  To locate the peak of the profile, we obtained 5 to 10
observations around the centers of the undeviated and deviated beams. Each
observation required manually rotating the turntable and reading the
turntable's position with a vernier scale.  We solved for the final
deviation angle $\delta$ by fitting Gaussians to the undeviated and
deviated profiles, and then calculating the angular difference between the
centers of the beams.

\subsection{Wavelengths and Temperatures}

	We measured the refractive index of \hbox{S-FTM16} at
17 wavelengths between 0.546 \micron\ and 2.600 \micron\ and at 5
temperatures (62 K, 78 K $\times 2$, 87 K, 97 K, and 296 K).  A
second set of measurements at 296 K was obtained in air with the dewar
windows removed.  Table \ref{tab:data} summarizes the measured
wavelengths: column 1 lists those wavelengths measured with the Si
detector and column 2 lists those wavelengths measured with the HgCdTe
detector.

	We had planned to obtain additional measurements at
temperatures between 100 K and room temperature, but the refractometer
was unable to operate at these temperatures.  We measured the
undeviated and deviated beams two to four times at each wavelength and
temperature, yielding a raw data set of 3,246 individual observations
and a reduced data set of 120 refractive indices.  The measurements
were obtained over five days in November 2003.

\subsection{Errors}

	Our measurements contain both statistical and systematic errors.  
The systematic errors are much larger than the statistical errors, but the
systematic errors are mostly removable.  For example, we measure the wedge
of the dewar windows by comparing room temperature measurements with and
without the dewar windows in place.  Similarly, we measure the focus
offset between the Si and HgCdTe detectors by observing $\lambda=1.129$
and $\lambda=1.200$ \micron\ with both detectors.  However, when we
compare the Ohara melt indices with our room temperature measurements (see
Figure \ref{fig:oc3}), we find a residual systematic offset at the
$\pm10\times10^{-5}$ level that varies with wavelength.  This systematic
offset is likely due to mechanical error in the turntable.  The
Soci\'{e}t\'{e} Instruments Physique PI-4 is specified to have $5\arcsec$
accuracy, but the turntable we used was relatively old and the
experimenter (W.\ Brown) noticed that the rotary handle had variable
resistance in different areas.  This suggests that the rotary mechanism
has lost its original accuracy.  We can fit and remove the systematic
table error, but in doing so we would introduce additional uncertainty in
the final index measurement.

	We can avoid systematic errors altogether by focusing on the
$\Delta n/\Delta T$ measurement of S-FTM16.  The window wedge, detector
offset, and rotary table error are independent of temperature, and so in
principle cancel out when calculating $\Delta n / \Delta T$.  The error in
individual $\Delta n / \Delta T$ values is thus dominated by the
statistical error in the index measurements.

	Most of the sources of statistical error in the index measurements
are angular uncertainties in the experimental apparatus that transform to
uncertainties in the index $n$ via Eqn.\ \ref{eqn:n}.  Table
\ref{tab:error} summarizes these errors.

\centerline{ \includegraphics[width=3.0in]{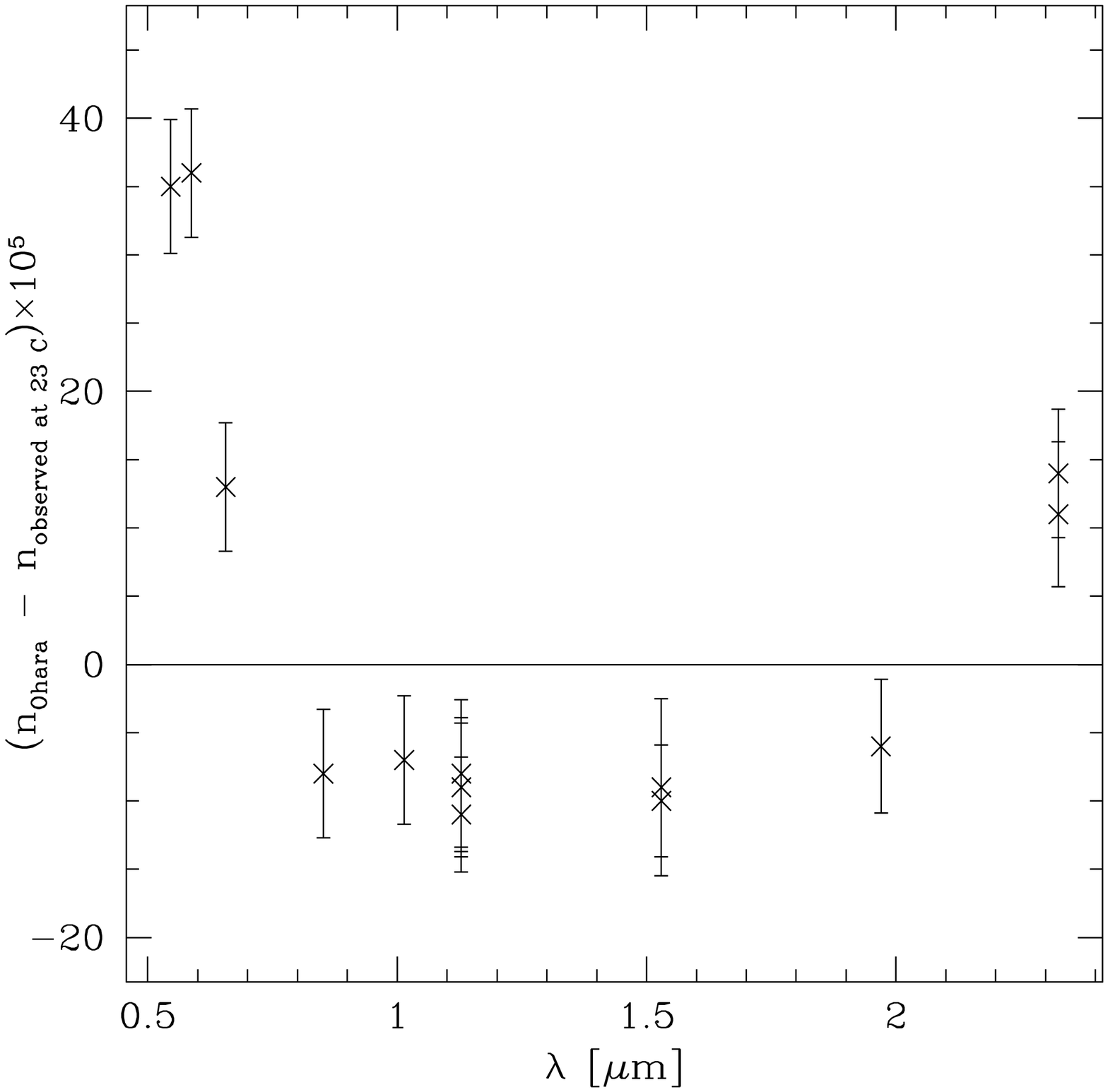} }
 \figcaption{ \label{fig:oc3}
	Difference between the Ohara melt values and our room temperature
index measurements.  The shape of the offset is the same at all
temperatures, and is likely due to mechanical error in the turntable.}

\subsubsection{Deviation Angle}

	We estimate the error in the deviation angle measurement from the
residuals of the Gaussian profile fits.  The residuals of the Gaussian
profile fits are typically $\pm4.1\arcsec$ for the Si detector and
$\pm5.1\arcsec$ for the HgCdTe detector.  Summing the undeviated and
deviated beam uncertainties in quadrature, the uncertainty in the
deviation angle contributes errors of $\pm2.1\times10^{-5}$ and
$\pm2.8\times10^{-5}$ to $n$ for the Si and HgCdTe detectors,
respectively.

\subsubsection{Apex Angle}

	We measured the apex angle $\alpha$ of the \hbox{S-FTM16} prism on
a Wild 79 spectrometer.  The average of ten measurements was
$\alpha=34^{\circ} 59\arcmin 38\arcsec\pm3\arcsec$.  This is statistically
identical to Ohara Corporation's measurement of the apex angle,
$\alpha_{\rm Ohara}=34^{\circ} 59\arcmin 39\arcsec\pm3\arcsec$.  The
$\pm3\arcsec$ uncertainty in the apex angle contributes an error of
$\pm2.2\times10^{-5}$ to $n$, and has a small wavelength dependence.

\subsubsection{Autocollimation}

	The deviation angle measurement assumes that the prism is aligned
with the undeviated collimated beam of light.  The autocollimation of the
prism is set by aligning the retro-reflected beam from the prism face onto
the exit slit of the Jarrell Ash monochrometer.  The retro-reflected beam
can be set with an uncertainty of $\pm0.5\arcsec$.  We maintain the
autocollimation by shining a He-Ne laser onto the prism face and making
sure the reflected spot on the wall ($\sim$8 m away) does not move.  If
the prism shifts (i.e.\ due to changes in temperature) we tip/tilt the
dewar to compensate.  The laser spot center can be maintained to
$\pm0.75\arcsec$.  Summing these uncertainties in quadrature, the
autocollimation of the prism contributes an error of $\pm0.3\times10^{-5}$
to $n$.

\subsubsection{Wavelength}

	We calibrated the wavelength readings on the Jarrell Ash
spectrometer with higher orders of a He-Ne alignment laser.  A linear fit
with wavelength leaves excessive residuals, and so our final calibration
uses a cubic fit.  The residuals of the fit are $\pm0.20$ nm.  In
addition, the dial on the spectrometer was read with a precision of
$\pm0.28$ nm. Summing these uncertainties in quadrature, the total error
in the wavelengths is $\pm0.34$ nm.  Propagating this through the Schott
equation shows that the wavelength uncertainty contributes an error of
$\pm0.5\times10^{-5}$ to $n$.

\subsubsection{Temperature}

	We average the temperature readings of the diodes above and below
the prism to obtain the prism temperature.  Because the observed
temperature gradient across the prism is always less than $1 ^{\circ}$C at
equilibrium, we conservatively estimate the uncertainty in the prism
temperature to be $\pm0.5 ^{\circ}$C.  Propagating this through a
$dn/dT$ relation shows that the temperature uncertainty contributes an
error of $\pm0.2\times10^{-5}$ to $n$.

\section{RESULTS AND DISCUSSION} \label{sec:results}

\subsection{Room Temperature $n$}

	The $\pm1\times10^{-5}$ accuracy of the Ohara melt information is
superior to our measurements, and so we report Ohara's melt values for
\hbox{S-FTM16's} room temperature 

\noindent \includegraphics[width=3.55in]{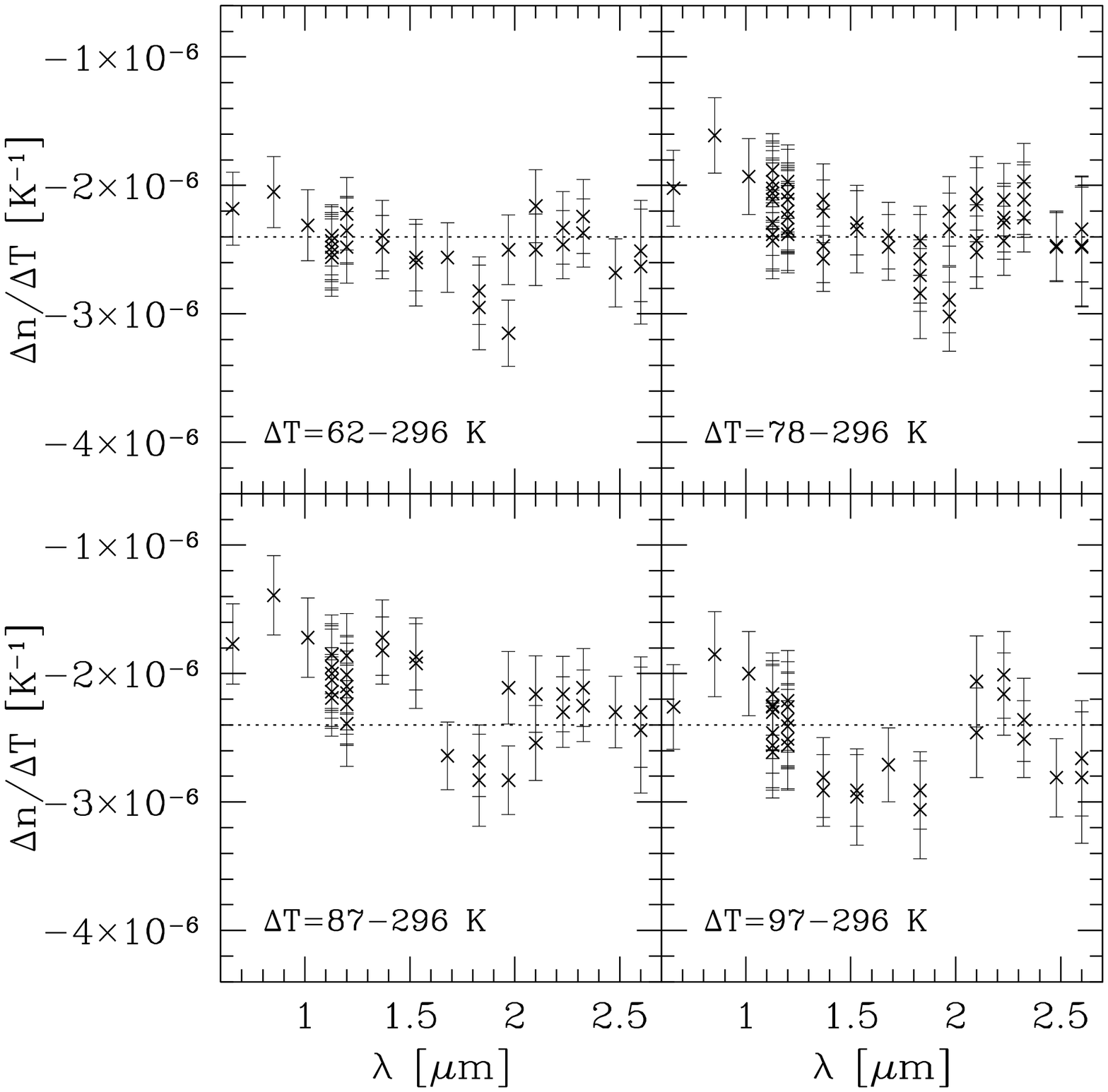}
 \figcaption{ \label{fig:dndt}
	Values of $\Delta n/\Delta T$ calculated between our four
cryogenic temperatures and room temperature.  Dotted lines show the
average value $\Delta n/\Delta T=-2.4\times10^{-6} \pm0.3\times10^{-6}$
K$^{-1}$ for all temperature differences.}

	~

\noindent index relative to air (Table
\ref{tab:ohara}). We fit the Ohara melt values with the six coefficient
Schott formula,
	\begin{equation} \label{eqn:schott} n^2 = A0 + A1 \lambda^2 + A2
\lambda^{-2} + A3 \lambda^{-4} + A4 \lambda^{-6} + A5 \lambda^{-8},
	\end{equation} where $\lambda$ is in \micron.  The best-fit Schott
coefficients are presented in Table \ref{tab:zemax}.  The residuals of the
best-fit Schott function are $\pm0.55\times10^{-5}$ in $n$.  It is
important to obtain melt information because \hbox{S-FTM16's} refractive
index will vary from melt to melt.

	Because this paper is primarily concerned with absolute indices,
Table \ref{tab:zemax} also presents the best-fit Schott coefficients for
\hbox{S-FTM16} in vacuum.  We correct the Ohara indices from air to vacuum
by multiplying the indices by the refractive index of air
$n(\lambda)_{25,760}$.  The refractive index of air at 25 $^{\circ}$C, 760
torr, and in the wavelength range 1 to 2.6 \micron\ is
$n(\lambda)_{25,760} \simeq 1.00026$ \citep{filippenko82}, with a
well-determined wavelength dependence at the $10^{-7}$ level.  The
air-to-vacuum transformation adds negligible error to the values of $n$.

\subsection{Cryogenic $n$ and $dn/dT$}

	We determine \hbox{S-FTM16's} absolute cryogenic refractive index
by calculating the difference between our cryogenic and room temperature
vacuum measurements.  Calculating the difference in index minimizes the
effects of systematic errors and any uncertainty in the calibration of our
data.  Our series of absolute index measurements thus reduces to a series
of absolute $\Delta n/\Delta T$ values between $\sim$77 K and room
temperature.  We look for wavelength-dependent terms in the $\Delta
n/\Delta T$ values, but, because of the relatively coarse
$\pm3.5\times10^{-5}$ precision of our indices, we do not find wavelength-
or temperature-dependent terms statistically significant.  Our vacuum
indices provide a direct measurement of absolute $\Delta n/\Delta T$.

	Figure \ref{fig:dndt} plots the values of absolute $\Delta
n/\Delta T$ between our four cryogenic temperatures and room temperature.
Measurements are plotted against wavelength.  We have two data sets at
room temperature that we use as comparison.  The room temperature
measurements in air have been corrected to vacuum.  The grand average of
all the values, shown by the dotted line in Figure \ref{fig:dndt}, is
$\Delta n_{abs}/\Delta T = -2.4\times10^{-6} \pm 0.3\times10^{-6}$
K$^{-1}$.

	We see apparent wavelength-dependent structure in the $\Delta
n/\Delta T$ measurements, but we determine that this wavelength-dependent
structure is not significant with respect to a constant fit.  Figure
\ref{fig:meanwl} plots all of our $\Delta n/\Delta T$ measurements
together.  Physically, we expect $\Delta n/\Delta T$ to have wavelength
dependence, but we expect the wavelength dependence to be small at our
temperatures and near-infrared wavelengths.  When we test high order
wavelength fits to the $\Delta n/\Delta T$ data set, we find that the
reduced $\chi^2$ and the RMS residuals do not significantly improve with
respect to a constant $\Delta n/\Delta T$.  For example, the linear fit in
Figure \ref{fig:meanwl} reduces the RMS residuals by a mere 6\%.  
Furthermore, the linear fit deviates from a constant $\Delta n/\Delta T$
by $\pm0.1\times10^{-6}$ K$^{-1}$ (a factor of three smaller than our
uncertainty) within the wavelength range 1 \micron\ to 2.6 \micron.  We
plot values of $\Delta n/\Delta T$ averaged over temperature (lower panel,
Figure \ref{fig:meanwl})  to further explore the possibility of
wavelength-depen-

\centerline{ \includegraphics[width=3.0in]{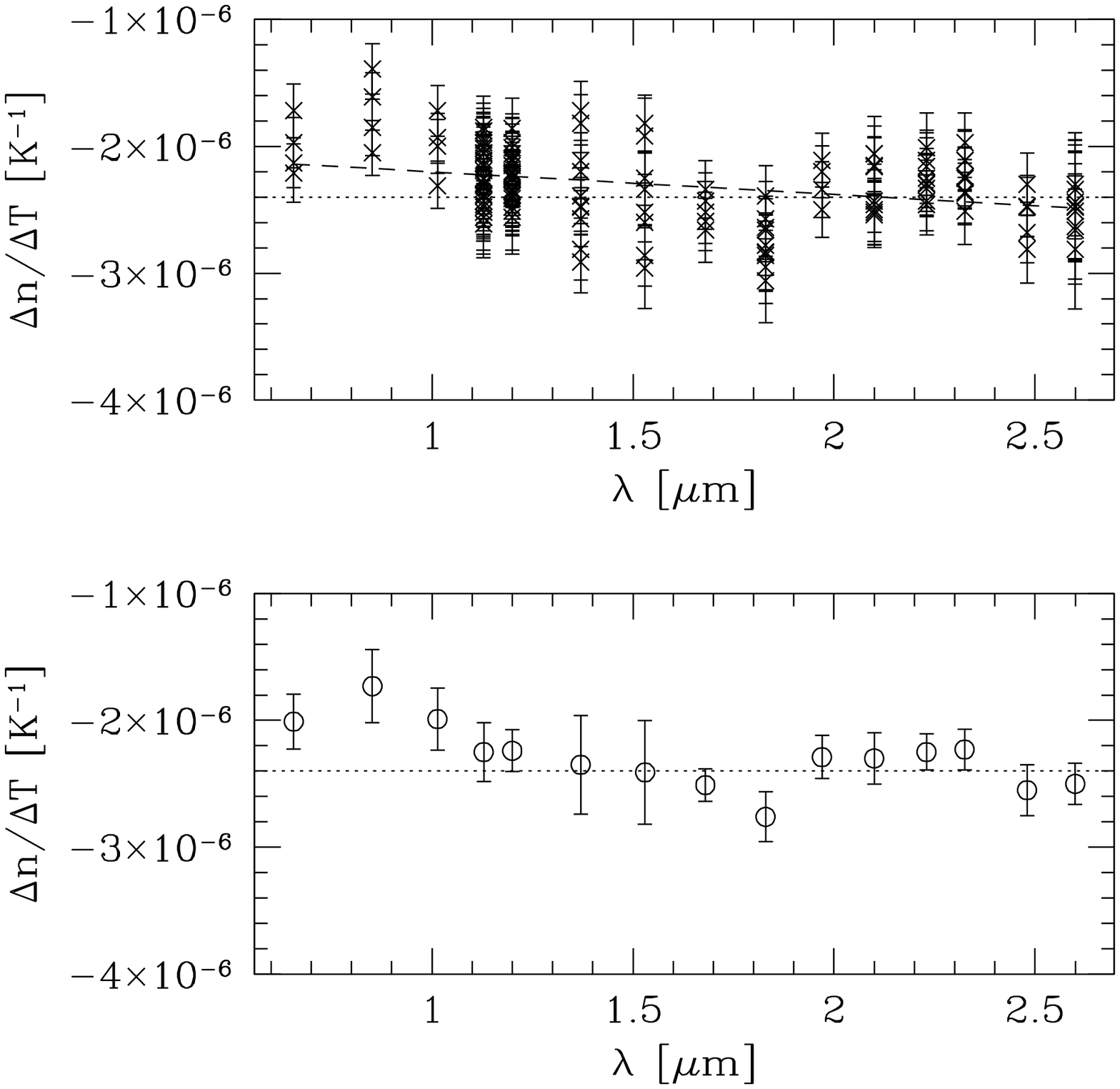}}
 \figcaption{ \label{fig:meanwl}
	The upper panel shows all $\Delta n/\Delta T$ measurements plotted
together; the dashed line is the linear best fit to the data and does not
significantly improve the residuals.  The lower panel shows the $\Delta
n/\Delta T$ values averaged over temperature, with the error bars showing
the RMS.  The horizontal dotted lines show our final fit $\Delta n/\Delta
T=-2.4\times10^{-6} \pm0.3\times10^{-6}$ K$^{-1}$.}

	~

\centerline{ \includegraphics[width=3.0in]{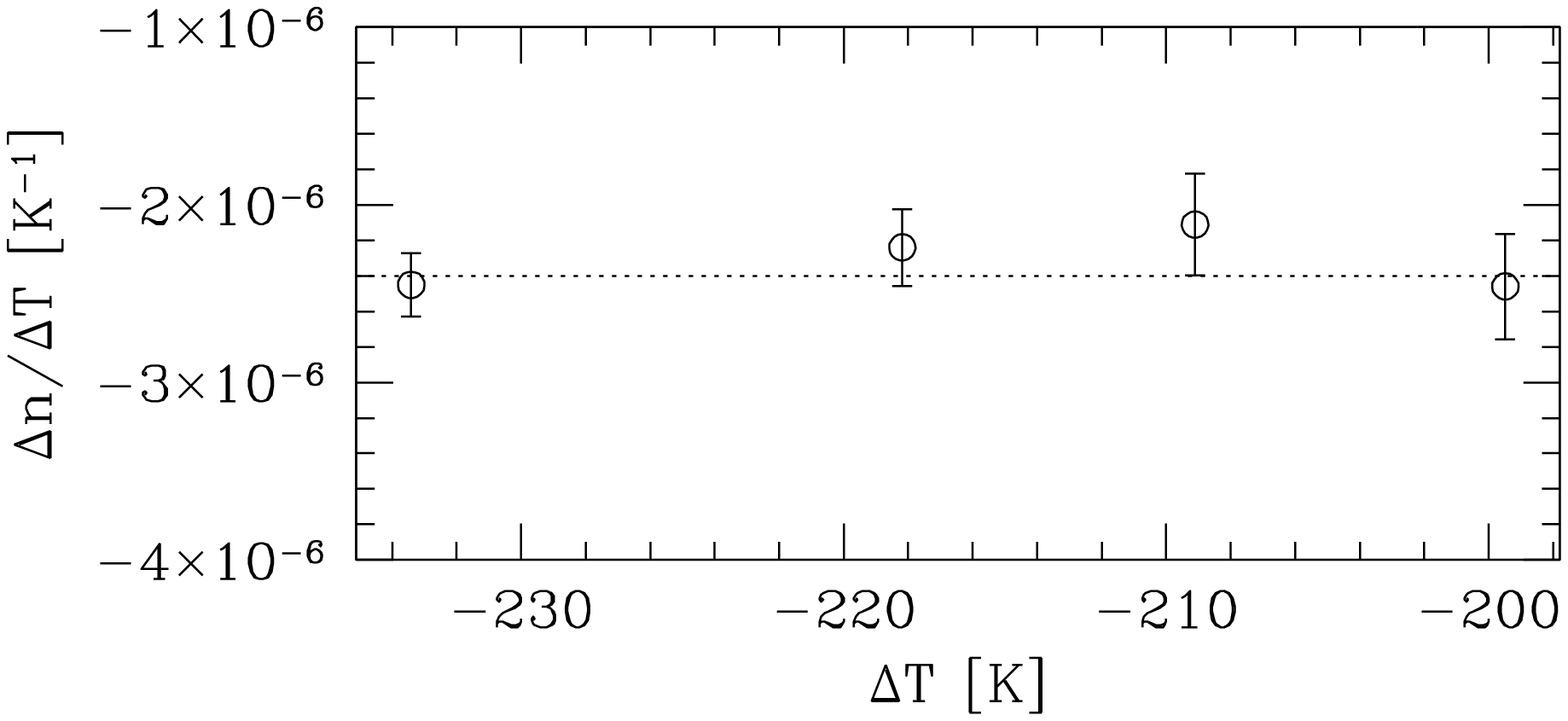}}
 \figcaption{ \label{fig:meant}
	Values of $\Delta n/\Delta T$ averaged over wavelength.  Error
bars show the RMS of the $\Delta n/\Delta T$ values at that temperature.  
The dotted line shows our final fit $\Delta n/\Delta T=-2.4\times10^{-6}
\pm0.3\times10^{-6}$ K$^{-1}$.}

\noindent \includegraphics[width=3.5in]{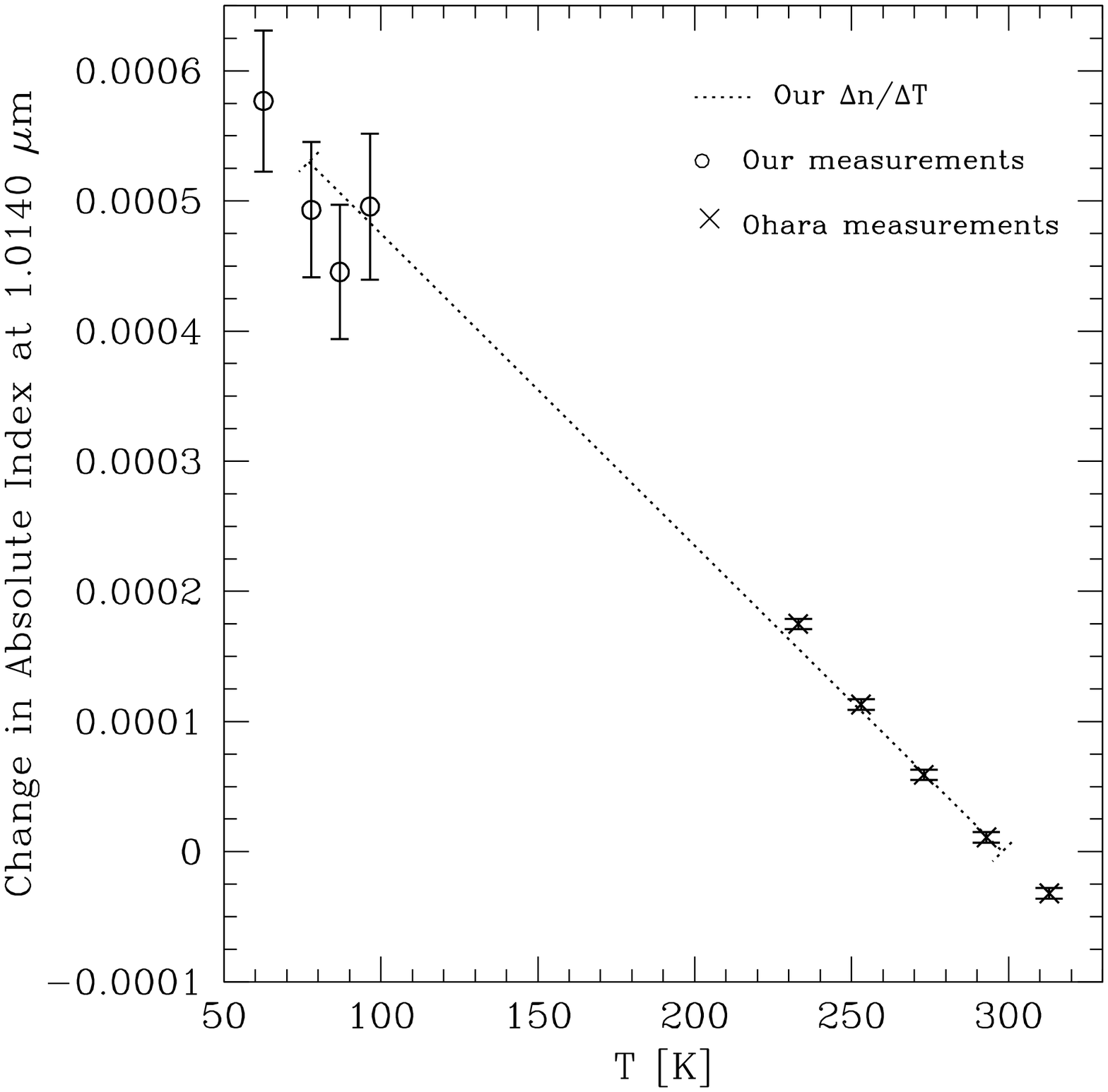}
 \figcaption{ \label{fig:dndt6a}
	A comparison of our absolute $\Delta n/\Delta T$ value and Ohara
absolute $\Delta n/\Delta T$ values integrated over appropriate ranges in
temperature.  The change in absolute index is normalized to 0 at 298 K.  
The dotted line shows our final fit $\Delta n_{abs}/\Delta T =
-2.4\times10^{-6} \pm 0.3\times10^{-6}$ K$^{-1}$ between 298 K and 77 K,
which nicely matches the Ohara $\Delta n/\Delta T$ values near room
temperature.}

	~

\noindent
dent structure.  The error bars show the RMS of the
$\Delta n/\Delta T$ values at that wavelength.  Only one of the 15 points
differs by 2 $\sigma$ from a constant $\Delta n/\Delta T$; eleven points
agree to better than 1 $\sigma$ with a constant $\Delta n/\Delta T$.  
Thus the data appears well described by a constant $\Delta n/\Delta T$.  
Because the error on the mean is negligible, we conclude that the
$\pm0.3\times10^{-6}$ K$^{-1}$ uncertainty is dominated by residual
systematic errors in the measurements.

	We now investigate temperature dependence in the $\Delta n/\Delta
T$ measurements.  Figure \ref{fig:meant} plots the $\Delta n/\Delta T$
values averaged over wavelength.  Error bars show the RMS for the $\Delta
n/\Delta T$ values at that temperature.  Our cryogenic measurements span
62 K to 97 K, and we see no obvious trend in this range of temperatures.  
Thus $\Delta n/\Delta T$ is well described by a constant value in the
temperature range 62 K to 97 K.


	We compare with Ohara catalog $dn/dT$ values as a check of our
measurements.  Note that we measure a single, integrated $\Delta n/\Delta
T$ value between room temperature and liquid nitrogen temperatures while
Ohara, on the other hand, measures $dn/dT$ in small temperature and
wavelength ranges.  A comparison can be made only if we integrate the
$dn/dT$ values over temperature and compare the changes in index, $\Delta
n$.  Figure \ref{fig:dndt6a} shows the change in absolute index at 1.0140
\micron, normalized to 0 at 298 K, for Ohara's integrated $dn/dT$ values
and our $\Delta n/\Delta T$ measurements (from Figure \ref{fig:meant}).  
The dotted line in Figure \ref{fig:dndt6a} shows our $\Delta
n_{abs}/\Delta T = -2.4\times10^{-6} \pm 0.3\times10^{-6}$ K$^{-1}$
between 298 K and 77 K.  Our $\Delta n/\Delta T$ nicely matchs the Ohara
$\Delta n/\Delta T$ values near room temperature.

	The cryogenic index of \hbox{S-FTM16} in vacuum is found by adding
the $\Delta n$ for the desired $\Delta T$.  Going from room 

\noindent \includegraphics[width=3.6in]{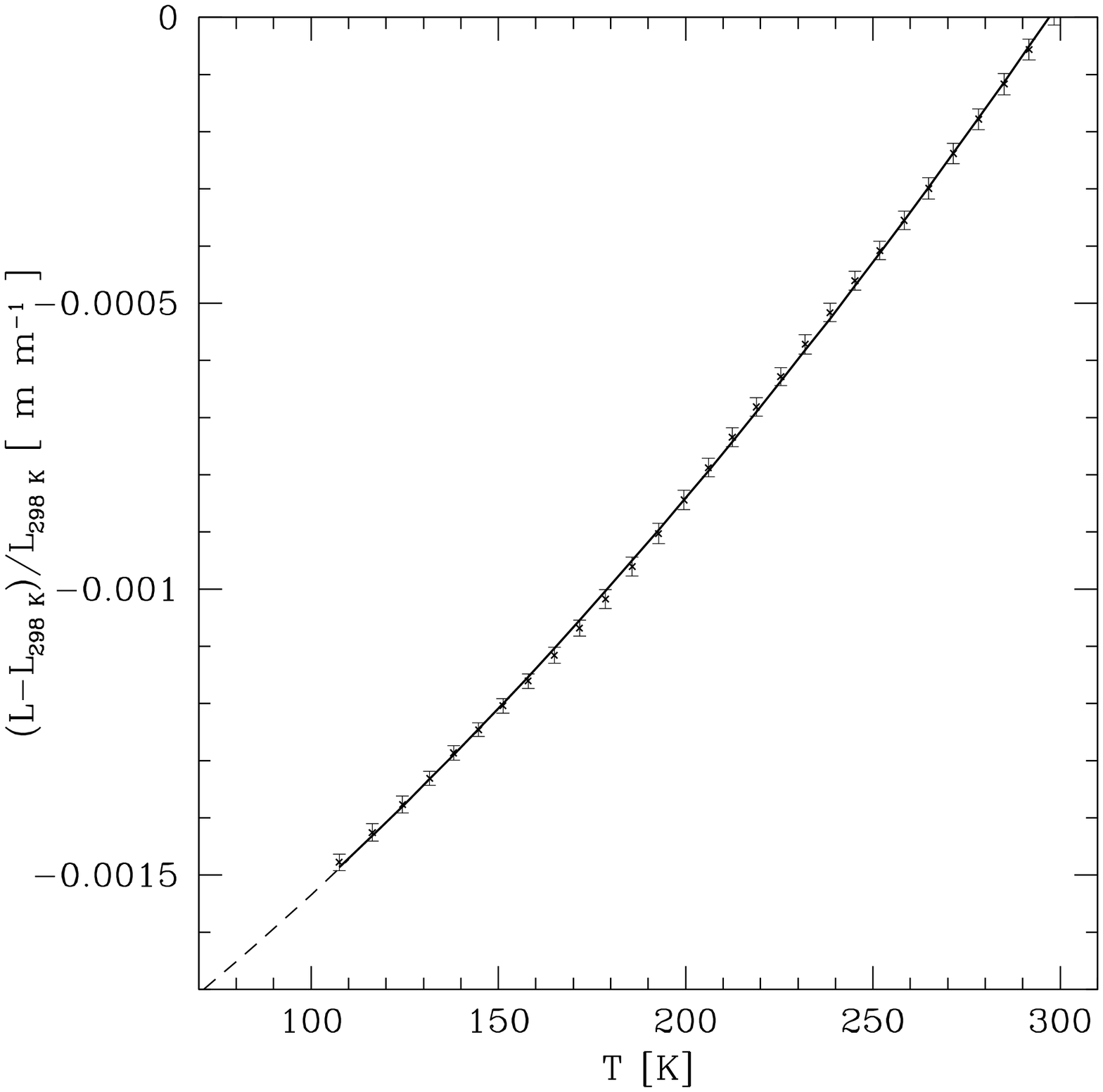}
 \figcaption{ \label{fig:dlt}
	 Integrated linear thermal expansion of \hbox{S-FTM16} with
respect to 298 K.  The solid line shows the best-fit quadratic, and the
dashed line shows the extrapolation of this fit to 70 K.}

	~

\noindent temperature
298 K to liquid nitrogen temperature 77 K, the absolute $\Delta n_{298
\rightarrow 77 {\rm K}} = +53\times10^{-5} \pm 7\times10^{-5}$ (see Figure
\ref{fig:dndt6a}).

	Finally, we investigate the effect of the 10\% uncertainty in
$\Delta n/\Delta T$ on a high-performance optical design.  The MMT
Magellan InfraRed Spectrograph (MMIRS) optical design, for example, uses
two \hbox{S-FTM16} elements in the camera \citep{mcleod04}.  Changing
\hbox{S-FTM16} in MMIRS by $\Delta n = \pm10\times10^{-5}$ results in a
small 0.3\% change in RMS spot size and a $\pm25$ \micron\ re-focus.  
Thus our 10\% determination of $\Delta n/\Delta T$ is more than adequate
to maintain the high performance of the MMIRS optical design.

\subsection{Cryogenic Coefficient of Thermal Expansion}

	We obtain cryogenic CTE measurements so that properly athermalized
lens mounts can be designed for \hbox{S-FTM16}.  On 13 April 2004 Harrop
Industries, Inc, performed a thermal dilatometric analysis on a sample of
\hbox{S-FTM16} cut from the same block as our prism.  The \hbox{S-FTM16}
sample was first cooled to 100 K (colder temperatures were not possible
with the apparatus), and then the sample's length was measured with a
dilatometer as it slowly warmed up.  The result is a direct measurement of
\hbox{S-FTM16's} integrated linear thermal expansion between 100 K and 298
K.

	Figure \ref{fig:dlt} shows \hbox{S-FTM16's} integrated linear
thermal expansion with respect to 298 K.  There is a clear second-order
term in the data, and the solid line shows the best-fit quadratic
	\begin{equation} (L-L_{298 {\rm K}})/L_{298 {\rm K}} =
-2.05\times10^{-3} + 4.27\times10^{-6} T + 8.84\times10^{-9} T^2,
\end{equation} where $L$ is the length at temperature $T$ (in K), and
$L_{298 {\rm K}}$ is the length at 298 K.  The dashed line in Figure
\ref{fig:dlt} shows the extrapolation of this fit to 70 K.  The integrated
linear 

\noindent \includegraphics[width=3.6in]{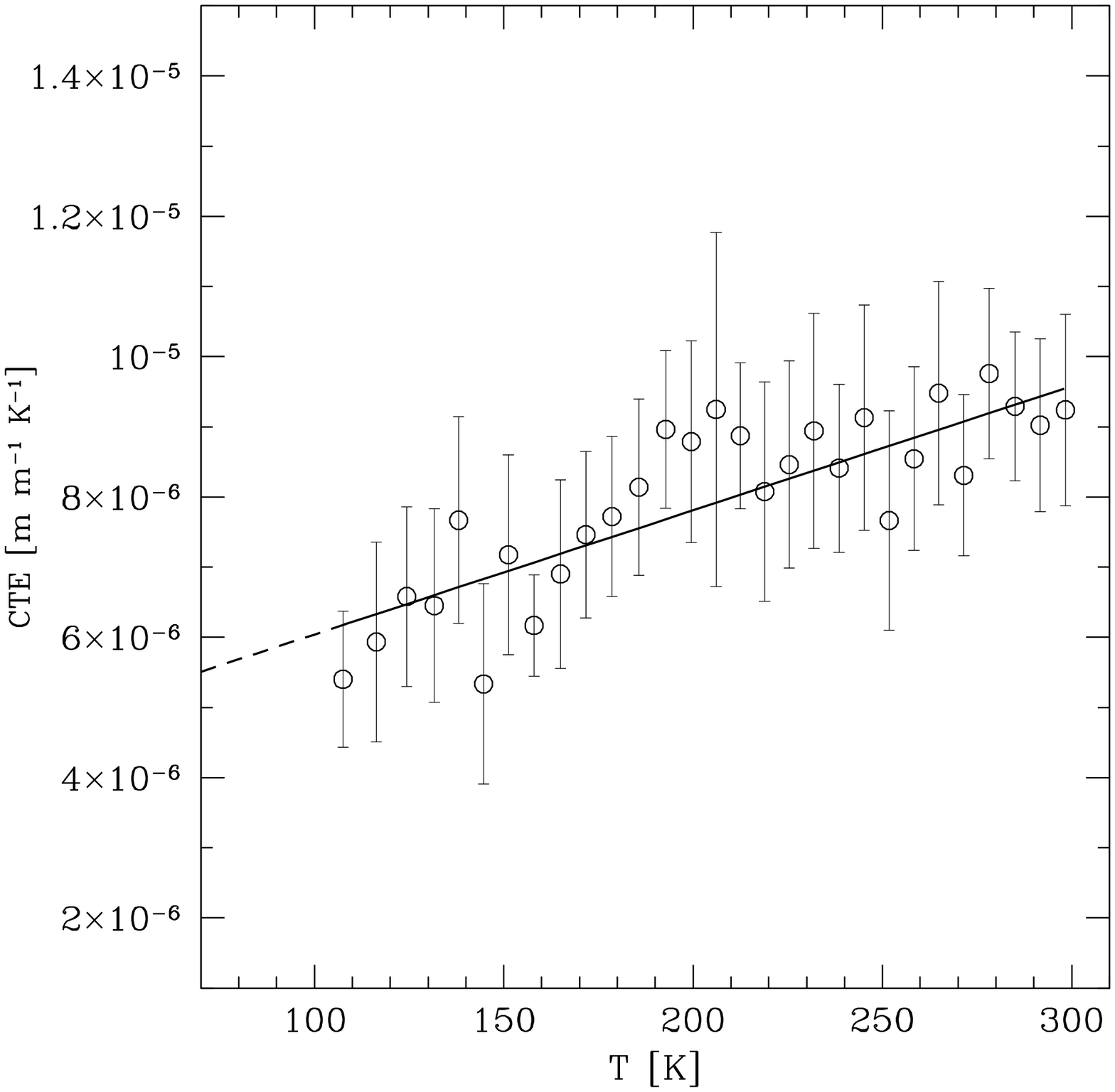}
 \figcaption{ \label{fig:cte}
	\hbox{S-FTM16's} linear coefficient of thermal expansion with
temperature.  The solid line shows the best-fit line, and the dashed line
shows the extrapolation of this fit to 70 K.}

	~

\noindent thermal expansion of \hbox{S-FTM16} between 298 K and 77 K is
-0.00167 $\pm 0.00001$ \hbox{m m$^{-1}$}.

	Figure \ref{fig:cte} shows \hbox{S-FTM16's} CTE with temperature.  
The CTE is simply the slope of the integrated linear thermal expansion
line at a given temperature.  The CTE values in Figure \ref{fig:cte} are
calculated by binning 16 dilatometer measurements together in
approximately 9 K intervals; the error bars show the RMS of the
measurements in a given bin.  The solid line drawn through the CTE values
in Figure \ref{fig:cte} is given by
	\begin{equation} {\rm CTE} = 4.27\times10^{-6} + 1.77\times10^{-8}
T, \end{equation} where CTE has units of \hbox{m m$^{-1}$} K$^{-1}$ and
$T$ has units of K.  The dashed line in Figure \ref{fig:cte} shows the
extrapolation of this fit to 70 K.  We find that \hbox{S-FTM16's} CTE is a
smoothly varying, positive quantity between 100 K and 298 K, consistent
with our $\Delta n/\Delta T$.  Ohara reports a CTE of $9\times10^{-6}$
\hbox{m m$^{-1}$} K$^{-1}$ between 240 K and 340 K, which is in good
agreement with our CTE measurements.

\section{CONCLUSIONS}\label{sec:conclusions}

	The glass \hbox{S-FTM16} is an important addition to the list of
optical materials for near-infrared instruments.  The glass has $>95\%$
internal transmission out to 2.4 \micron, and is currently the only
intermediate dispersion material available to near-infrared optical
designs. High-performance optical designs require accurate knowledge of
material properties, especially at the cryogenic temperatures at which
near-infrared instruments operate.  Thus we have undertaken cryogenic
refractive index measurements for \hbox{S-FTM16}.

	Our measurements were made with the Cryogenic Refractometer at the
University of Arizona.  The experimental apparatus uses the modified
minimum deviation method to measure refractive index.  Individual
measurements have $\pm3.5\times10^{-5}$ precision in $n$.  We use Ohara's
high accuracy melt information to define \hbox{S-FTM16's} relative
refractive index at room temperature.  We convert the relative indices to
absolute indices.  We then use our vacuum measurements to calculate the
change in \hbox{S-FTM16's} refractive index between room temperature and
77 K, independent of systematic errors.

	We find $\Delta n_{abs}/\Delta T = -2.4\times10^{-6}
\pm0.3\times10^{-6}$ K$^{-1}$, valid in the wavelength range 0.6 \micron\
to 2.6 \micron\ and between room temperature and liquid nitrogen
temperatures.  This corresponds to an absolute $\Delta n_{298 \rightarrow
77 {\rm K}} = +53\times10^{-5} \pm 7\times10^{-5}$ between 298 K and 77 K.  
We find no statistical evidence for wavelength- or temperature-dependent
terms to the uncertainty of our measurements.  The 10\% accuracy of our
$\Delta n/\Delta T_{abs}$ determination is more than adequate to maintain
the high-performance of the MMIRS optical design. The integrated linear
thermal expansion of \hbox{S-FTM16} between 298 K and 77 K is -0.00167
$\pm 0.00001$ \hbox{m m$^{-1}$}.

\acknowledgements

	We are grateful to the late Richard J.\ Elston for his interest
and support.  Richard provided the S-FTM16 glass samples as part of the
FLAMINGOS-2 project.  We also thank Dr.\ Jim Palmer for allowing us to use
the Cryogenic Refractometer, and for his assistance with set-up and
measurements.  Funding for this project was provided in part by W.\
Brown's Harvard-Smithsonian CfA Fellowship.  The cryogenic CTE
measurements were paid by the MMIRS project, supported by AURA through the
National Science Foundation under AURA Cooperative Agreement AST 0132798,
as amended.


\def\arraystretch{1.1}

\begin{deluxetable}{lcc}	
 \tablewidth{0pt}
 \tablecolumns{3}
 \tablecaption{Ohara S-FTM16 Indices\tablenotemark{a}\label{tab:ohara}}
 \tablehead{  \colhead{$\lambda$ (\micron)}
 & \colhead{$n_{rel}$} & \colhead{$\sigma_n\times 10^5$} }
	\startdata
0.435835 & 1.61528 & 1.0 \\
0.486133 & 1.60529 & 1.0 \\
0.546075 & 1.59737 & 1.0 \\
0.587562 & 1.59338 & 1.0 \\
0.63280  & 1.58997 & 1.0 \\
0.656273 & 1.58847 & 1.0 \\
0.85211  & 1.58028 & 1.0 \\
1.01398  & 1.57636 & 1.0 \\
1.12864  & 1.57423 & 1.0 \\
1.52958  & 1.56830 & 1.0 \\
1.97009  & 1.56215 & 1.0 \\
2.32542  & 1.55674 & 1.0 \\
	\enddata
 \tablenotetext{a}{At standard air $(T,P)$ = (\hbox{25 $^{\circ}$C}, 760
torr).} \end{deluxetable}

\begin{deluxetable}{ccccccc}	
\tablewidth{0pt}
\tablecolumns{7}
\tablecaption{Ohara Values of S-FTM16's Absolute $dn/dT$\label{tab:oharadndt}}
\tablehead{
 \colhead{Temp ($^{\circ}$C)} &
	\multicolumn{6}{c}{$dn_{abs}/dT$ ($10^{-6}/ ^{\circ}$C)} \\
 \colhead{} & \colhead{$\lambda 0.4358$} & \colhead{$\lambda 0.4800$} & 
 \colhead{$\lambda 0.5461$} & \colhead{$\lambda 0.5893$} & 
 \colhead{$\lambda 0.6328$} & \colhead{$\lambda 1.0140$} \\
}
\startdata
$-30$ &  $-0.8$ &  $-1.5$ &  $-2.2$ &  $-2.4$ &  $-2.6$ &  $-3.1$ \\
$-10$ &  $-0.3$ &  $-1.1$ &  $-1.7$ &  $-1.8$ &  $-2.1$ &  $-2.7$ \\
~ 10  &  ~ 0.2  &  $-0.6$ &  $-1.3$ &  $-1.5$ &  $-1.8$ &  $-2.4$ \\
~ 30  &  ~ 0.7  &  $-0.2$ &  $-0.9$ &  $-1.2$ &  $-1.5$ &  $-2.1$ \\
~ 50  &  ~ 1.0  &  ~ 0.2  &  $-0.6$ &  $-0.9$ &  $-1.2$ &  $-1.8$ \\
~ 70  &  ~ 1.4  &  ~ 0.5  &  $-0.4$ &  $-0.7$ &  $-1.0$ &  $-1.6$ \\
	\enddata
\end{deluxetable}

\begin{deluxetable}{cc}	
\tablewidth{0pt}
\tablecolumns{2}
\tablecaption{Wavelengths Measured\label{tab:data}}
\tablehead{
  \colhead{$\lambda$(Si)} &
  \colhead{$\lambda$(HgCdTe)} \\
  \colhead{\micron} & \colhead{\micron}
}
\startdata
0.546	& 1.129	\\
0.588	& 1.200	\\
0.656	& 1.370	\\
0.852	& 1.530	\\
1.014	& 1.680	\\
1.129	& 1.830	\\
1.200	& 1.970	\\
	& 2.100	\\
	& 2.230	\\
	& 2.325	\\
	& 2.480	\\
	& 2.600	\\
	\enddata
	\end{deluxetable}

\begin{deluxetable}{lcc}	
\tablewidth{0pt}
\tablecolumns{3}
\tablecaption{Summary of Statistical Errors\label{tab:error}}
\tablehead{ \colhead{Error Source} & \colhead{Si} & \colhead{HgCdTe} \\
& \colhead{$\sigma_n \times 10^5$} & \colhead{$\sigma_n \times 10^5$}
}
\startdata
Deviation angle measurement	& 2.1 & 2.8 \\
Apex angle measurement		& 2.2 & 2.1 \\
Autocollimation alignment	& 0.3 & 0.3 \\
Wavelength setting		& 0.5 & 0.5 \\
Temperature uncertainty 	& 0.2 & 0.2 \\
				& ==  & == \\
TOTAL ERROR			& 3.1 & 3.6 \\
	\enddata
\end{deluxetable}

\begin{deluxetable}{cll} 	
\tablewidth{0pt}
\tablecolumns{3}
\renewcommand{\arraystretch}{1.1}
\tablecaption{S-FTM16 Schott Coefficients\label{tab:zemax}}
\tablehead{ \colhead{Schott coef}
 & \colhead{$(T,P)=(+25~ ^{\circ}$C, 760 torr)}
 & \colhead{$(T,P)=(+25~ ^{\circ}$C, 0 torr)}
}
\startdata
A0 & $\phm{-} 2.47395             $ & $\phm{-} 2.47522             $ \\
A1 & $       -1.00689\times10^{-2}$ & $       -1.00713\times10^{-2}$ \\
A2 & $\phm{-} 2.10265\times10^{-2}$ & $\phm{-} 2.11039\times10^{-2}$ \\
A3 & $\phm{-} 9.30325\times10^{-4}$ & $\phm{-} 8.93809\times10^{-4}$ \\
A4 & $       -3.30921\times10^{-5}$ & $       -2.44143\times10^{-5}$ \\
A5 & $\phm{-} 7.08359\times10^{-6}$ & $\phm{-} 6.41271\times10^{-6}$ \\
	\enddata
\end{deluxetable}

\end{document}